# Ultralow and Selective Infrared Emission from MXenes

Meikang Han, Danzhen Zhang, Christopher Eugene Shuck, Yury Gogotsi*

*A. J. Drexel Nanomaterials Institute and Department of Materials Science and Engineering, Drexel University, Philadelphia, PA 19104, USA*

*Correspondence to Prof. Yury Gogotsi (Y. G.), Email: gogotsi@drexel.edu

**ABSTRACT**

Low and selective infrared emission materials are necessary for next generation thermal technologies, including passive heating, infrared identification, and photothermal conversion. Here, we report on the intrinsic infrared radiation properties of three MXenes: $Ti_3C_2T_x$, $Ti_3CNT_x$, and $V_4C_3T_x$. The infrared emissivity of 200 nm thick $Ti_3C_2T_x$ coating is lower than 0.06 between wavelengths of 3-27 μm, while $Ti_3CNT_x$ and $V_4C_3T_x$ have values of 0.13 and 0.26, respectively. Investigation using a laboratory-based passive heating process showed that MXene coatings suppress the thermal radiation. The infrared identification capability was also validated using a patterned MXene fabric. This work paves the way for developing MXene-based smart wearable devices and garments capable of selective and localized thermal management.
KEYWORDS: MXene, infrared emissivity, thermal radiation, coating, patterned fabric

**Introduction**

Modulating thermal radiation in the infrared (IR) range is critical for next generation thermal technologies, such as selective heating/cooling, IR camouflage, thermal circuits, smart textiles, and others.[1-4] Currently, the state of the art in adaptive radiation heating/cooling uses metamaterials to manipulate the phonon behavior,[5] including topological transitions, photonic crystals, epsilon-near-zero surfaces, and others.[6-8] However, utilization of these precise engineered structures limits their applications in highly integrated devices and large-scale thermal manipulation components due to cost and scalability issues. Alternative approaches, such as active electrical modulation of thermal emissivity, are promising, but have limited range and rely on the inherent thermal properties of the structures.[9, 10] For passive heating management, low IR emission/absorption is required to suppress energy dissipation by Stefan-Boltzmann's law. Conventional polished metals (Al, Ag, Ni, etc.) have ultralow IR emissivity (< 0.1), but are difficult to produce; typically, thin metal coatings/films are deposited using magnetron sputtering or electroless plating, especially for the complex architectures.[11, 12] Therefore, it is still challenging to achieve selective thermal radiation using passive intrinsic IR emission material characteristics.

Two-dimensional transition metal carbides and nitrides (MXenes) have shown diverse light-matter interactions in a very broad electromagnetic frequency range,[13, 14] owing to their anisotropic electronic and optical properties, active surfaces, and a high density of states at the Fermi level.[15-19] For instance, MXenes present active plasmon resonances from the ultraviolet (UV) to near-IR range as a function of structure and composition.[20] 16 different MXene compositions were used for electromagnetic interference (EMI) shielding in gigahertz frequency range and the performance was adjusted by controlling the M-site chemistry.[21, 22] There are a few reports on light-to-heat conversion of $Ti_3C_2T_x$ MXene for thermal management.[23-26] However, little is known about the intrinsic thermal radiation properties of MXenes, especially in the IR range. Furthermore, more than 30 kinds of stoichiometric MXenes have been experimentally synthesized, with numerous MXene compositions theoretically predicted.[19, 27-29] It is likely that MXenes will have diverse IR radiation properties depending on the composition, atomic arrangement, and surface chemistry. Moreover, due to the solution processability of MXenes, thin MXene films/coatings from nanometer- to micrometer-thickness can be readily manufactured using multiple techniques.[30-32] The ease of processing coupled with the natural hydrophilicity and flexibility of MXenes facilitates the deposition of MXenes on a wide variety of surfaces, enabling the rapid treatment of surfaces for thermal management.

Herein, we report preliminary data on the low IR emissivity of three MXene ($Ti_3C_2T_x$, $Ti_3CNT_x$, and $V_4C_3T_x$) coatings for object identification/camouflage and other thermal management applications. MXene-coated glass with different coating thickness was fabricated by a simple spray-coating process. The IR emissivity of different MXene coatings was measured using a Fourier-transform infrared (FTIR) spectrometer based on Kirchhoff's law of thermal radiation. The IR radiation performance was validated through IR imaging of MXene coatings on a hot plate. As a proof of concept, we also fabricated a $Ti_3C_2T_x$-patterned cotton fabric for image identification, demonstrating the low thermal radiation of MXene. This work illustrates the potential use of MXenes in films, garments and other systems for selective and localized thermal management.

**Results and Discussion**

Three different MXenes were synthesized and delaminated into few-layer flakes, as shown in Figure 1a-c. $Ti_3C_2T_x$, $Ti_3CNT_x$, and $V_4C_3T_x$ flakes had a lateral size of > 3 μm, > 2 μm, and > 300 nm, respectively. The complete etching and delamination was further confirmed by X-ray diffraction (XRD) patterns of MXene films (Figure 1d). The (002) peaks of $Ti_3C_2T_x$ and $Ti_3CNT_x$ are at 7.2° (2θ) while $V_4C_3T_x$ is at 5.3°, indicating that the $V_4C_3T_x$ film had a larger $d$-spacing (16.6 Å). As shown in Raman spectra (Figure 1e), the peaks at 201 cm$^{-1}$ and 718 cm$^{-1}$ for $Ti_3C_2T_x$ correspond to $A_{1g}$ (Ti, O, C) and $A_{1g}$ (C) vibrations.[33] $Ti_3CNT_x$ shows a similar spectrum with $Ti_3C_2T_x$, but with a weaker $A_{1g}$ (Ti, O, C) vibration (~201 cm$^{-1}$) and the absence of peak at 718 cm$^{-1}$ due to the random arrangement of C and N atoms. $V_4C_3T_x$ shows $A_{1g}$ vibrations at ~264 cm$^{-1}$ and ~660 cm$^{-1}$ and $E_g$ vibration at ~480-540 cm$^{-1}$.

To investigate the IR emissivity of different MXenes, ~200 nm thick coating of each MXene was fabricated on glass substrates by spray-coating. As shown in Figure 2a, each MXene coating is different visually and under optical microscope, and is commensurate with previous reports.[20] The emissivity ($\varepsilon$)/absorbance ($A$) spectra of MXene coatings were calculated using the measured reflectance ($R$) with a FTIR spectrometer over a wavelength range of 0.3-27 μm, as shown in Figure 2b. The MXene coatings show low emissivity in mid-IR range (> 2.5 μm) and a nonlinear trend in the UV-vis-NIR range (0.3-2.5 μm). The 200 nm thick $Ti_3C_2T_x$ coating has an emissivity of < 0.1 over the entire mid-IR range, while $Ti_3CNT_x$ and $V_4C_3T_x$ coatings are below 0.2 and 0.4, respectively. Particularly, in the atmospheric window of thermal radiation (e.g. 8-14 μm), $Ti_3C_2T_x$, $Ti_3CNT_x$, and $V_4C_3T_x$ have an average emissivity of 0.06, 0.13, and 0.26, respectively, implying that MXenes have the capability to suppress thermal radiation. Considering diversity of MXenes in both elemental composition and structure, along with the already known differences in optical and electronic properties, it is probable that other MXenes can achieve selective infrared emission with transparent coatings. However, additional studies are needed to understand the IR radiation mechanism of MXenes.

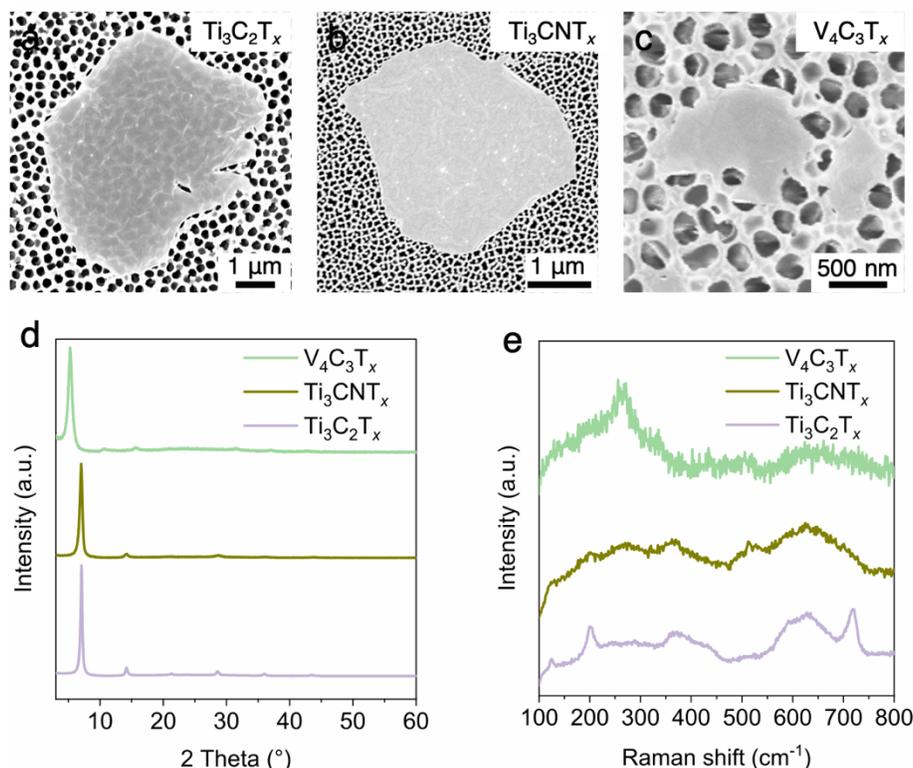

**Figure 1.** Scanning electron microscopy (SEM) images of three MXene flakes: (a) $Ti_3C_2T_x$, (b) $Ti_3CNT_x$, and (c) $V_4C_3T_x$. (d) XRD patterns and (e) Raman spectra of different MXene coatings.

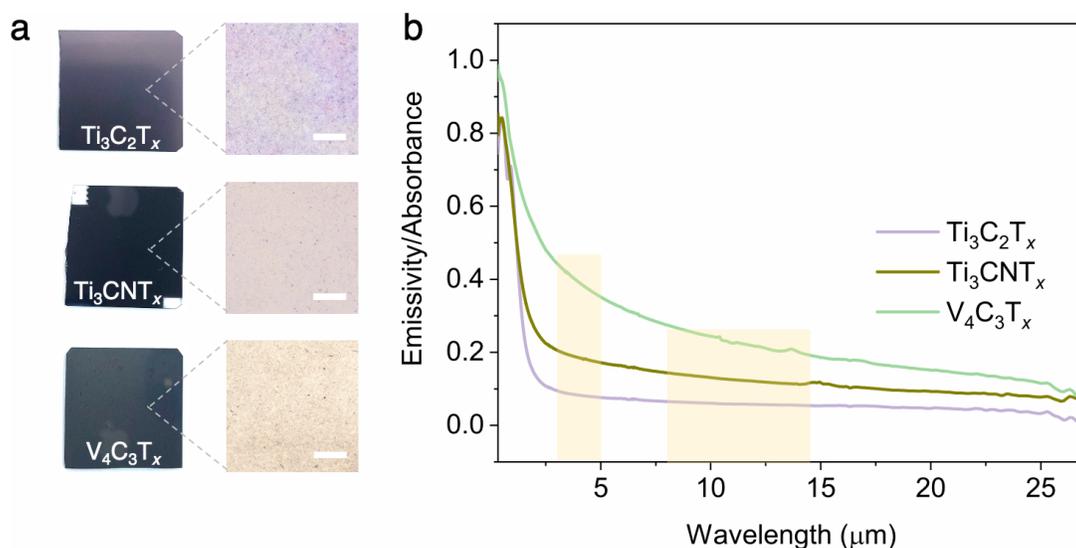

**Figure 2**. (a) The digital (left) and optical (right) images of MXenes ($Ti_3C_2T_x$, $Ti_3CNT_x$, and $V_4C_3T_x$) spray-coated on glass substrates. (b) The emissivity/absorbance of ~200 nm thick MXene coatings in a wavelength range of 0.3-27 μm. The scale bar in optical images is 40 μm.

To validate the low IR emission of MXenes, the MXene-coated glass slides were placed on a hot plate at 110 °C for 300 s with the apparent temperature measured with an IR camera. As shown in Figure 3a, for all three MXenes, the temperature increased rapidly in the first 30 s, and then stabilized after ~300 s. After stabilization, , temperatures of 39.5 °C and 44 °C were recorded for $Ti_3C_2T_x$ and $Ti_3CNT_x$, respectively, and attributed to their ultralow IR emission. Because $V_4C_3T_x$ has a higher emissivity, it exhibits an apparent steady-state temperature of 66.3 °C after 300 s. While higher than the Ti-based MXenes, this detected value is still lower than the real surface temperature of 110 °C, indicating the capability of all tested MXenes to suppress thermal radiation. From the IR images of MXene coatings on a hot plate in different time (Figure 3c), it was visually observed that the temperature changes of different MXene coatings with the increasing time. $Ti_3C_2T_x$ and $Ti_3CNT_x$ have a similar color change while $V_4C_3T_x$ shows larger temperature change from 57.0 to 66.3 °C. We further investigated the thermal radiation behavior of $Ti_3C_2T_x$ using different coating thicknesses (~200, 600, and 1300 nm). As shown in Figure 3b, the steady-state temperature of the 1300 nm thick $Ti_3C_2T_x$ coating is 38.3 °C while the 200 nm coating reaches 39.2 °C. In addition, there is no significant difference for the three samples in the first 30 s which is the rapid heating process. This indicates that the nanometer-thick $Ti_3C_2T_x$ coating has a prominent IR identification performance, which is attributed to its ultralow IR emissivity rather than low thermal conductivity.

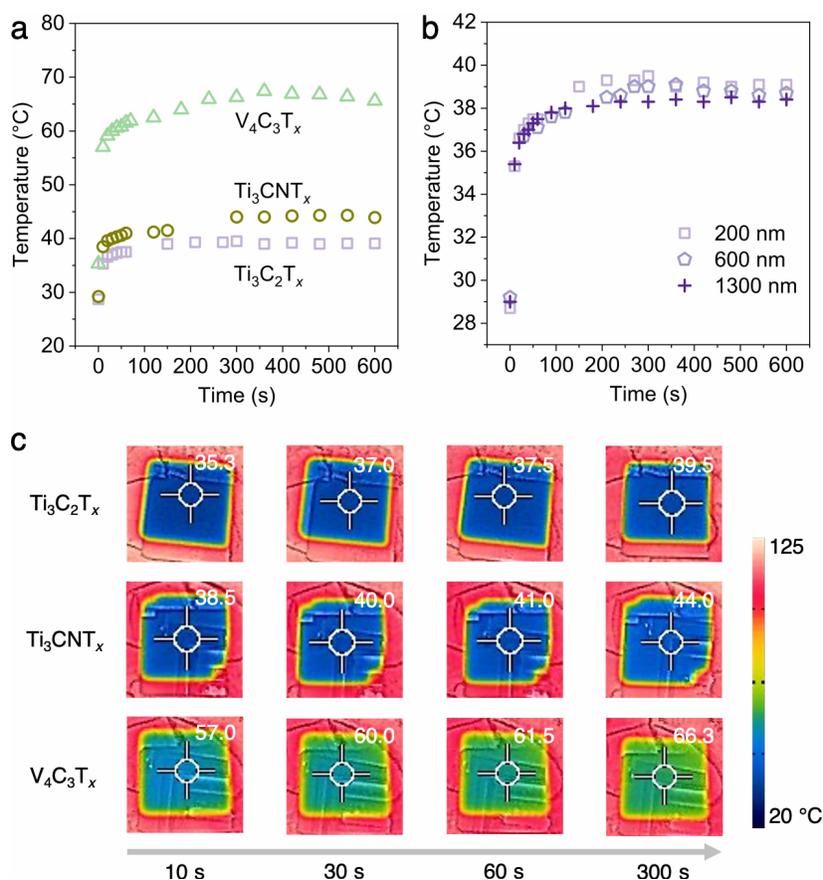

**Figure 3**. Temperature as a function of time for (a) different MXene coatings ($Ti_3C_2T_x$, $Ti_3CNT_x$, and $V_4C_3T_x$) with a thickness of ~200 nm, and (b) $Ti_3C_2T_x$ coatings with different thicknesses (~200 nm, ~600 nm, and ~1300 nm) on a hot plate at 110 °C. (c) Infrared images of different MXene coatings at different time.

A Ti$_3$C$_2$T$_x$ dragon-shaped pattern was sprayed onto a cotton fabric to demonstrate a proof-of-concept application (Figure 4a). When the fabric was worn, the patterned region only reads ~28 °C using the IR camera, which is lower than the uncoated region (~32 °C), as shown in Figure 4b. This demonstrates that a Ti$_3$C$_2$T$_x$ coating, even when on a rough, porous, or uneven surfaces, still has a lower IR emissivity. This result indicates that MXene may potentially be used for IR camouflage coatings, IR identification tags, and localized thermal management systems. Compared to traditional metals with low IR emission, MXenes can be more easily integrated with fabrics and complex devices owing to their solution processability. In addition to IR emission properties, we have demonstrated that MXene-coated fabrics have high-performance EMI shielding in gigahertz ranges.[34] It is promising to develop smart MXene-coated garments with multifunctional capabilities, including IR stealth/identification, EMI shielding, and wireless communication.

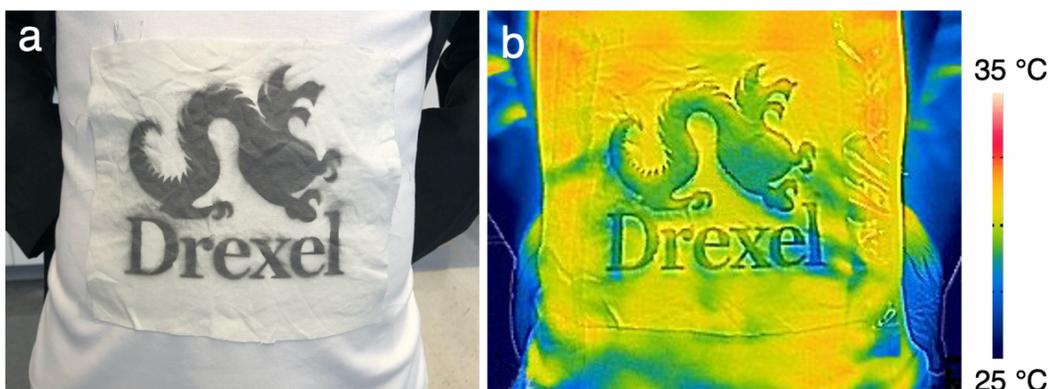

**Figure 4**. (a) Visible and (b) infrared images of a Ti$_3$C$_2$T$_x$-coated fabric attached on a human body, showing the low infrared emission of Ti$_3$C$_2$T$_x$.

**Conclusions**

In summary, we have demonstrated that MXenes with different compositions (Ti$_3$C$_2$T$_x$, Ti$_3$CNT$_x$, and V$_4$C$_3$T$_x$) have ultralow IR emissivity in the mid-IR range, and show non-linear optical absorbance in UV-vis-NIR range. ~200 nm thick Ti$_3$C$_2$T$_x$, Ti$_3$CNT$_x$, and V$_4$C$_3$T$_x$ films have an average IR emissivity of 0.06, 0.13, and 0.26, respectively, in the atmospheric window. MXene-coated glass shows lower IR-detected temperatures than the heating object. IR identification was achieved in a Ti$_3$C$_2$T$_x$-patterned cotton fabric. This work clearly shows the potential of MXenes for the integration into smart wearables/devices with selective IR identification/stealth, thermal insulation, and localized thermal manipulation with different MXenes.

**Methods**

*Materials*

Ti (99.5%, −325 mesh, Alfa Aesar), Al (99.5%, −325 mesh, Alfa Aesar), TiC (99.5%, ~2 μm, Alfa Aesar), AlN (98%, 10 μm, Aldrich), V (99.5%, −325 mesh, Alfa Aesar), C (graphite, 99%, −325 mesh, Alfa Aesar), HF (48.5−51%, Acros Organics), HCl (36.5−38%, Fisher

Chemical), LiCl (99%, Acros Organics), and tetramethylammonium hydroxide (25 wt%, Acros Organics).

*Synthesis of MAX phases*

For $Ti_3AlC_2$, the precursor TiC:Ti:Al powders were mixed in an atomic ratio of 2:1.25:2.2, following previous reports. The powders were ball milled at 70 rpm for 18 h in polyethylene bottles with zirconia balls (2:1 mass ratio of balls to powder). The powders were then placed into an alumina crucible, which was placed into a high temperature furnace (Carbolite Gero) and heated at 1400 °C for 2 h. The samples were heated and cooled at a rate of 3 °C/min with Ar continually flowing at 200 ccm. The sintered compact was drilled with a TiN coated bit, then sieved to below 32 μm. The powders were washed with 9 M HCl for 24 h to remove residual intermetallics and metals, then were repeatedly washed with DI water. For $Ti_3AlCN$ (and $V_4AlC_3$) a similar procedure was followed, except the precursors Ti:AlN:C (V:Al:C) were mixed in a 3:1:1 (4:1.5:3) atomic ratio, and the sample was heated at 1500 °C for 2 h.

*Synthesis of $Ti_3C_2T_x$ and $Ti_3CNT_x$*

$Ti_3C_2T_x$ and $Ti_3CNT_x$ were synthesized by the selective etching of the corresponding MAX phase powders ($Ti_3AlC_2$ and $Ti_3AlCN$; -325 mesh) with an acids mixture (HF and HCl). The synthesis procedure was reported in our previous work.[21, 35] Typically, 2 mL of HF, 12 mL of HCl, and 6 mL of deionized (DI) water were mixed firstly. After that, 1 g of MAX phase powder was added to the solution and stirred for 24 h at room temperature. After etching, the reacted solution was washed with DI water using the centrifuge at 3500 rpm for 2 mins until pH > 6. The obtained sediment was mixed with 1 g of LiCl in 50 mL of DI water. The mixture was shaken for 15 mins, and then centrifuged at 3500 rpm for 10 mins several times until the sediment was swelled. The swelled sediment was dispersed in DI water and then centrifuged at 7500 rpm for 3 mins. After that, the black supernatant was collected for spray-coating.

*Synthesis of $V_4C_3T_x$*

$V_4C_3T_x$ were synthesized by the selective etching of $V_4AlC_3$ with HF. 1 g of $V_4AlC_3$ powder was added into 20 mL of HF and stirred at 35 °C for 8 days. After etching, the reacted solution was washed with DI water through centrifugation (3500 rpm, 2 min) until pH > 6. After that, the washed sediment was added into a solution with 1 g of TMAOH and 10 mL of DI water and stirred for 24 h. The mixture was centrifuged several times with DI water at 10000 rpm for 10 mins until pH < 8. At last, the solution was centrifuged at 3500 rpm for 10 min. The supernatant was the dispersion of delaminated $V_4C_3T_x$ flakes in water.

*Fabrication of MXene coatings*

MXene coatings was fabricated by spray-coating from MXene aqueous solution. Microscopic glass slide (Fisher Scientific) was used as the substate. Before spray-coating, the slide was cleaned by ethanol with sonication. The cleaned glass slide was plasma-treated (Tergeo Plus, Pie Scientific) at 100 W with oxygen flow at 3 sccm for 5 mins. MXene colloidal solution with a concentration was sprayed onto the glass substrates manually. The sprayed samples were dried in a vacuum oven at 70 °C for 12 h. MXene-coated cotton fabric was fabricated using a similar process. Before spray-coating, the fabric was washed using a liquid detergent (Synthrapol®).

*Characterization*

The morphology of MXene flakes was observed using scanning electron microscopy (SEM; Zeiss Supra 50VP, Germany). A 3D laser scanning confocal microscopy (Keyence, VK-X1000, Japan) was used to observe the coating surface and measure the coating. XRD patterns of MXene coatings were measured using a Rigaku SmartLab (Tokyo, Japan) operating at 40 kV/30 mA with Cu K$\alpha$ radiation. Raman spectra were measured with an inverted reflection mode Renishaw (2008, Gloucestershire, UK) instrument, equipped with 63× (NA = 0.7) objectives and a diffraction-based room-temperature spectrometer. The laser line used was 633 nm and the laser power was kept around 0.1 mW. The infrared images and temperature were recorded using an IR camera (FLIR E8xt, USA). The camera emissivity was set as 0.94. The reflectivity ($R$) of MXene coatings in UV-vis-MIR range (0.3-27 μm) was measured using a FTIR spectrometer (Invenio-X, Bruker, Germany). The emissivity ($\varepsilon$) was calculated based on Kirchhoff law ($\varepsilon$=1-$R$).

## Acknowledgements

We thank Dr. Sergey Shilov (Bruker Scientific, *LLC*) for the emissivity measurement of MXene coatings. XRD and SEM analyses were performed using instruments in the Materials Characterization Core at Drexel University. We thank Geetha Valurouthu for the help with collecting Raman spectra and Kyle Matthews for the assistance with SEM observations.

## Conflict of Interest

The authors have filed for a patent on the infrared emission of MXenes.[36] The authors declare no other conflict of interest.